\documentclass[amsmath,pre,preprint,aps,showpacs]{revtex4}
\usepackage{graphicx}
\begin{document}
\title{On the structure and spectrum of classical two-dimensional clusters \\
with  a logarithmic interaction potential}
\author{B. Partoens}
\email{bart.partoens@ua.ac.be}
\affiliation{Departement Natuurkunde, Universiteit Antwerpen (Campus Drie Eiken) \\
Universiteitsplein 1, B-2610 Antwerpen, Belgium}
\author{P. Singha Deo}\email{deo@bose.res.in}  \affiliation{S. N. Bose National Centre for Basic
Sciences \\ JD Block, Sector - III \\ Salt Lake,  Kolkata 98, India}

\begin{abstract}
We present a numerical study of the effect of the repulsive logarithmic inter-particle
interaction on the ground state configuration and the frequency spectrum of a confined
classical two-dimensional cluster containing a finite number of particles. In the case
of a hard wall confinement all particles form one ring situated at the boundary of the
potential. For a general $r^n$ confinement potential, also inner rings can form and we
find that all frequencies lie below the frequency of a particular mode, namely the
breathing-like mode. An interesting situation arises for the parabolic confined system
(i.e. $n=2$). In this case the frequency of the breathing mode is independent of the
number of particles leading to an upper bound for all frequencies. All results can be
understood from Earnshaw's theorem in two dimensions. In order to check the sensitivity
of these results, the spectrum of vortices in a type II superconductor which, in the
limit of large penetration depths, interact through a logarithmic potential, is
investigated.
\end{abstract}

\maketitle

\section{Introduction}
In recent years, classical clusters with a finite number of particles moving in a
two-dimensional plane were studied for different kinds of confinement and interaction
potentials. Bedanov {\it et al.}~\cite{bedanov,schweigert} studied the classical system
with a finite number of particles interacting through a repulsive $1/r$ potential and
moving in a two-dimensional plane, confined by a parabolic potential. The $1/r$
potential is the Coulomb potential in a three-dimensional world. This theoretical
system models experimental realizations, such as electrons on the surface of liquid
helium~\cite{helium}, electrons in quantum dots in high magnetic fields~\cite{reimann},
colloidal suspensions~\cite{colloidal} and confined plasma crystals~\cite{plasma}. A
detailed investigation of the structure~\cite{bedanov} and the spectral
properties~\cite{schweigert} of these 2D clusters was carried out. Recently, ground
state and metastable configurations and/or spectral properties of 2D clusters with a
parabolic confinement potential but different interaction potentials were investigated
(see Ref.~\cite{bolton,bedanov,schweigert,minghuitransition,koulakov,lini} for a $1/r$
potential, Ref.~\cite{lini,lozovik} for a dipole interaction, Ref.~\cite{lini,lozovik}
for a logarithmic potential, Ref.~\cite{lini,candido} for screened Coulomb interaction,
...) and with a $1/r$ Coulomb interaction potential but with different confinement
potentials (see Ref.~\cite{farias} for a Coulomb confinement potential,
Ref.~\cite{minghui} for a hard wall potential).

Laughlin~\cite{laughlin} has shown that for strongly interacting electrons in a
magnetic field at certain fractional fillings $(1/3, 1/5, \ldots)$ of the Landau
levels, the maxima in the pair correlation function correspond to the equilibrium
positions in a classical one dimensional plasma where the electrons interact via a
repulsive logarithmic potential. The idea was extended to quantum dots~\cite{manninen}
with few electrons as well as bosons and to rings~\cite{viefers} with a few electrons.
It was shown that the electrons actually get localized at the positions where the pair
correlation function peaks and the excitations of the system can be understood as
rotational modes of the center of mass and vibrational modes. The eigenenergies of the
system can be easily determined from the frequencies of the classical normal modes to a
high degree of accuracy.

In this paper we consider systems with various confinement potentials containing
particles interacting through a repulsive \textit{logarithmic} interaction. As the
solution of the 2D Poisson equation yields a logarithmic potential, one can say that
the Coulomb potential in a 2D world is logarithmic. Vortices in a film of liquid Helium
interact through a logarithmic potential~\cite{vorthelium}. Also vortices in a type II
superconducting 2D film, for a low concentration of vortices, is expected to interact
through a logarithmic potential~\cite{sup}. The logarithmic interaction between
vortices was used to study the stable vortex configurations in a disk shaped
superconductor~\cite{supvort}. Since the vortices are infinitely long in the
$z$-direction, the symmetry makes it effectively 2D. Hence the model may have some
experimental realizations and it may be possible to verify experimentally the
predictions made here.

In Section II we describe our model systems. The results for the ground state
configurations are shown in Section III, and for the spectra in Section IV. Section V
tests the sensitivity of the results by looking at the spectrum of vortices in a type
II superconductor, which interact through a logarithmic potential for a large
penetration depth. Our conclusions are given in Section V.

\section{Model and numerical approach}

The Hamitonian of a 2D system of $N$ charged particles in a $r^n$ confinement potential
and interacting through a repulsive logarithmic potential is given by
\begin{equation}
H=\sum_{i=1}^N \frac{1}{2} m \omega_0^2 R^2 \left( \frac{r_i}{R}\right)^n +
\sum_{i>j}^{N}V(\left| \vec{r}_{i}-\vec{r}_{j}\right|),
\end{equation}
where $m$ is the mass of the particle, $\omega_0$ the radial confinement frequency, and
$\vec{r}_i=(x_i,y_i)$ the position of the $i$th particle with $r_i \equiv |\vec{r}_i|$.
A hard wall confinement is obtained for $n\rightarrow \infty$ and in this case $R$
equals the radius of the hard wall. The interaction potential is taken to be
logarithmic:
\begin{equation}
V(r)= -\beta\ln(r/R).
\end{equation}

We can write the Hamiltonian in a dimensionless form if we express the coordinates and
energy in the following units: $ r' = \beta^{1/n} \alpha^{-1/n} R^{(n-2)/n}, E' =
\beta$, with $\alpha=\frac{1}{2}m\omega_0^2$. The dimensionless Hamiltonian is given by
\begin{equation}
H = \sum_{i=1}^N r_i^n - \sum_{i>j}^{N}\ln \left| \vec{r}_{i}-\vec{r}_{j}\right|.
\end{equation}
In the limit of a hard wall confinement, the lengthunit becomes $r' \rightarrow R$.

We also present results for the actual interaction potential for vortices in a type II
superconductor for $\lambda\rightarrow\infty$ as well as for smaller values of
$\lambda$ (with $\lambda$ the penetration depth). Vortices in a type II superconductor
interact with a potential~\cite{tinkham}
\begin{equation}
V_{s}(r)= \beta K_0\left(\frac{r}{\lambda}\right), \label{vs}
\end{equation}
with $K_0$ the zero-order Hankel function with imaginary argument, $\beta =
\Phi_0/(2\pi\lambda^2)$ and $\Phi_0=hc/2e$ the flux quantum. In the limit of $\lambda
\rightarrow \infty$, this potential becomes logarithmic:
\begin{equation}
V_s(r) \rightarrow - \beta \ln {r} , \mbox{ for } \lambda \rightarrow \infty \mbox{ (up
to a constant)}.
\end{equation}
With the above units, the dimensionless form of this interaction potential is
\begin{equation}
V_s (r) = K_0\left(\frac{r}{\lambda}\right),
\end{equation}
where now also $\lambda$ is expressed in the unit $r'$.

The numerical method to obtain the ground state configuration is based on the Monte
Carlo simulation technique supplemented with the Newton method in order to increase the
accuracy of the energy. The latter technique is outlined and compared with the Monte
Carlo technique in Ref.~\cite{schweigert}. The eigenmode frequencies are obtained from
the eigenvalues of the dynamical matrix
\begin{equation}
E_{\alpha \beta ,ij}=\left.\frac{\partial ^{2}E}{\partial r_{\alpha ,i}\partial
r_{\beta ,j}}\right|_{r_{\alpha ,i}=r_{\alpha ,i}^{n}},
\end{equation}
where $\left\{ r_{\alpha ,i}^{n}\right\} $ is the ground state configuration. The
eigenvalues of the dynamical matrix are the squared eigenfrequencies of the system. The
eigenfrequencies are expressed in the unit $\omega' =\sqrt{ E'/r'^2/m}$.

\section{Ground state configurations}

\begin{figure}
\includegraphics[width=8.6cm]{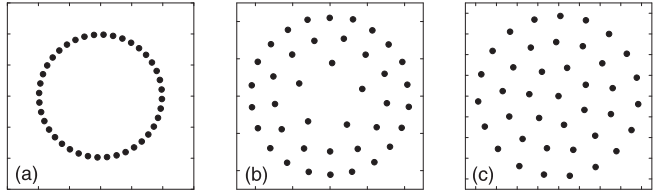} \caption{Ground state configurations for 40
particles interacting through a logarithmic potential for (a) a hard wall confinement,
(b) a confinement with $n=3$ and (c) for a parabolic confinement. The scale is
different in each figure, but the distance between the ticks is always one length
unit.} \label{groundstate}
\end{figure}

In this section we will discuss the numerically obtained ground state configurations
for different powers of the confinement potential and show how they can be understood
from Earnshaw's theorem. Fig.~\ref{groundstate} shows as an example the ground state
configurations for 40 particles for (a) a hard wall confinement (i.e.
$n\rightarrow\infty$), (b) a confinement with $n=3$ and finally (c) a parabolic
confinement (i.e. $n=2$). One can see that in the case of hard wall confinement all
particles are situated on one ring at the boundary of the potential. This is true for
any number of particles in the system, also for clusters with many particles. When the
confinement differs from the hard wall confinement, particles can be situated in the
central region. Notice that the density of particles for $n=3$ is much larger at the
edge than in the central region, while for the parabolic confinement the density is
uniform.
\begin{figure}
\includegraphics[width=8.2cm]{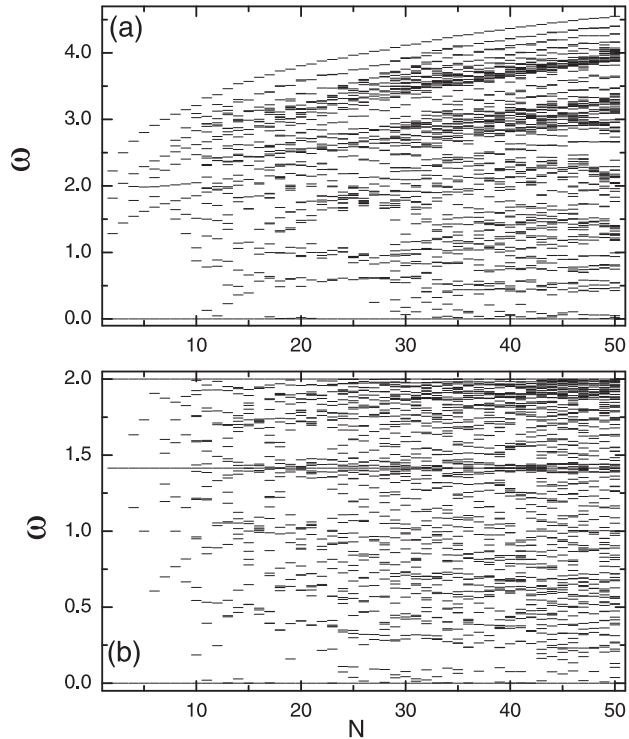}
\caption{The eigenfrequencies for 2 up to 50 particles interacting through a
logarithmic potential for (a) a $n=3$ confinement potential and (b) a parabolic
confinement potential.} \label{freqs}
\end{figure}
These observations can be understood from Earnshaw's theorem~\cite{earnshaw} in two
dimensions. Since in the present case the inter-particle potential is logarithmic we
can apply the 2D version of Gauss's theorem
\begin{equation}
\oint_C {\mathbf E} \cdot d{\mathbf r} \sim \int_S \rho({\mathbf r}) dS,
\end{equation}
where $S$ is the surface enclosed by $C$, and $\rho({\mathbf r})$ is the enclosed
charge. Now it is easy to show that charged particles on a 2D surface can not have a
stable static equilibrium when they interact through a logarithmic potential (i.e. the
2D Coulomb potential). If we take a small circle anywhere in the 2D region with no
enclosed charges, then, since the line integral of the field along the circle has to be
zero, the number of outgoing lines of force are equal to the number of incoming ones.
As long as there are outgoing lines of force, a charge left inside this circle can
lower its energy by moving along it. Thus in case of hard wall confinement, particles
can not be in a stable static equilibrium if they are not in contact with the hard wall
potential. Our simulations indeed show that all particles are pushed on one ring
situated at the boundary of the hard wall. This is different from the system with a
hard wall confinement in which the particles interact through a $1/r$ potential. In
Ref.~\cite{bedanov} it was shown that stable multiple ring structures can be found if a
sufficiently large number of particles are present in the 2D system.

As shown above the logarithmic interaction potential alone can not result in a stable
multi-ring configuration. It is only possible if the hard wall confinement is changed
into some soft wall $r^n$ confinement potential (as shown in Figs.~\ref{groundstate}(b)
and (c)). In this case, when a particle in the central region is moved in the outward
radial direction, the restoring force is given by the confinement potential alone. This
must balance the nonrestoring forces provided by the Coulomb repulsion with the other
particles. For the parabolic confinement case, this results in a uniform density as the
restoring force is everywhere the same.

\section{The eigenmode spectrum}
\begin{figure}
\includegraphics[width=8.6cm]{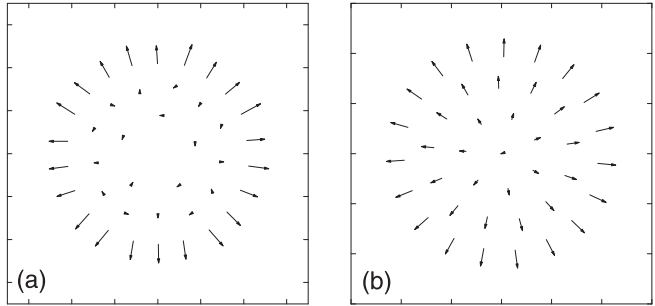} \caption{(a) The breathing-like mode for $N=40$
corresponding to the highest frequency for $n=3$. The breathing modes for $N=40$ in the
case of a parabolic confinement.} \label{modes}
\end{figure}

Also the eigenmode spectrum for systems with a logarithmic interaction potential
differs from all the previously studied cases. Again the results can be understood from
Earnshaw's theorem. The eigenfrequencies of the system with logarithmic interaction for
a $n=3$ confinement potential are, as an example for a soft wall confinement, shown in
Fig.~\ref{freqs}(a) for $N=2$ up to 50. Notice that there is a pronounced highest
frequency branch. We checked the corresponding eigenmodes and they all show the same
behaviour: all outer particles move in the radial direction and have a large amplitude.
An example of these breathing-like modes is shown in Fig.~\ref{modes}(a) for $N=40$. It
is interesting to look now at the eigenfrequencies in the case of a parabolic
confinement potential, which is shown in Fig.~\ref{freqs}(b). Again the mode
corresponding to the highest frequency is the breathing mode (shown in
Fig.~\ref{modes}(b) for $N=40$). But, as the frequency of the breathing mode in the
parabolic case is independent of the number of particles (and given by
$\omega_{\text{breathing}}=\omega_{\text{max}}=2$~\cite{thomson}), there is an upper
bound for the frequencies.

For confinement potentials with $n>2$ the potential is steeper than the parabolic
confinement at large distances from the center. So for the confinement with $n>2$ a
mode with larger velocities at the boundary than at the center overtakes the breathing
mode to be the highest mode. But in all these modes there is a simultaneous contraction
of a large number of particles so we can speak of breathing-like modes.

Normal modes are effectively a motion of free particles moving in a resultant potential
created by the other particles and the confinement potential. So whenever we try to
move a particle from its static equilibrium position, there are restoring forces that
bring it back. Then, addition of extra particles complicates the effective potential in
which the individual particles are moving and creates higher normal modes. This
potential can become steeper and steeper as more particles are added. This results in
higher and higher normal modes. But this argument breaks down for the logarithmic
interaction for which Gauss's theorem can be applied: as already shown in previous
section, the confinement potential is the only restoring force in the outward radial
direction and forms an upper bound to the restoring force. Addition of new particles
does not create a more complicated effective potential but it can only result in a
slight increase in the frequency of the breathing-like mode due to the increase of the
nonrestoring Coulomb force on the outer particles.

\section{Vortices in superconductors}
\begin{figure}
\includegraphics[width=8.6cm]{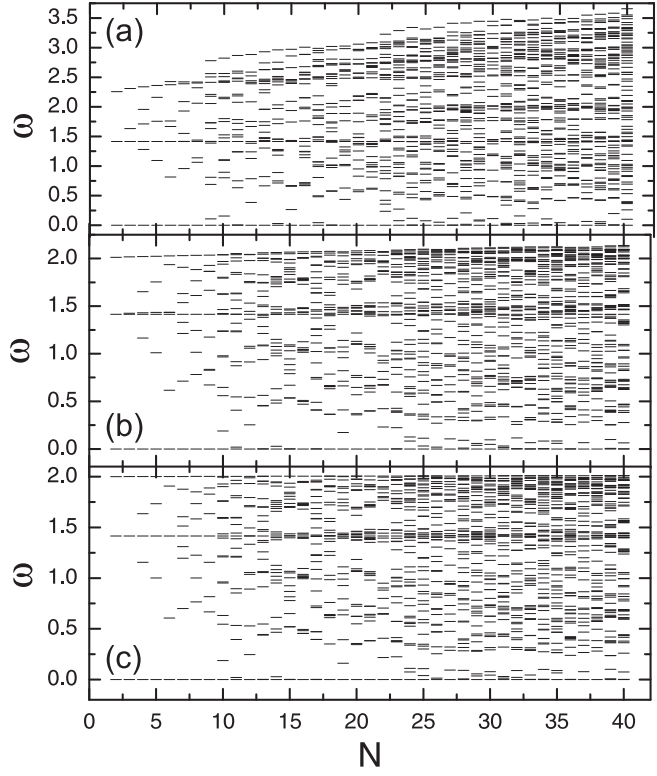} \caption{The spectrum for the parabollically confiend system interacting
through the potential $V_s(r)$ as a function of the number of particles $N$ for a)
$\lambda=1$, b) $\lambda=10$ and c) $\lambda=50$.} \label{freqsvs}
\end{figure}
\begin{figure}
\includegraphics[width=12.6cm]{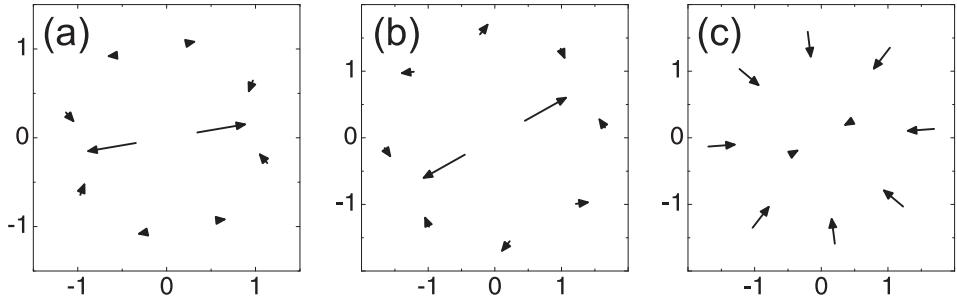} \caption{The eigenmodes corresponding to
highest frequency for $N=10$ for a) $\lambda=1$, b) $\lambda=10$ and c) $\lambda=50$.}
\label{modesvs10}
\end{figure}
\begin{figure}
\includegraphics[width=12.6cm]{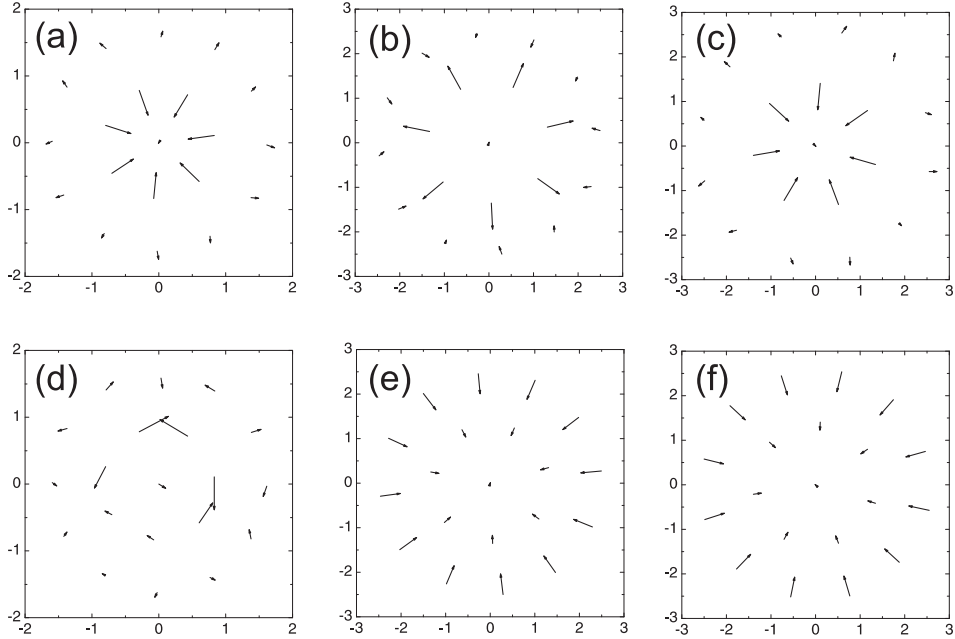} \caption{The eigenmodes corresponding to
highest frequency for $N=20$ for a) $\lambda=1$, b) $\lambda=10$ and c) $\lambda=50$,
and corresponding to the highest but one frequency for d) $\lambda=1$, e) $\lambda=10$
and f) $\lambda=50$.} \label{modesvs20}
\end{figure}

As mentioned in the Introduction, the logarithmic potential is sometimes a good model
for the interaction between vortices in a type II superconductor. In order to check how
important small deviations from this logarithmic potential are, we also studied the
spectrum of vortices in a type II superconductor for a more generally valid interaction
potential. In Ref.~\cite{tinkham} it is shown that the interaction potential between
vortices in a type II superconductor is given by Eq.~(\ref{vs}) if the penetration
length $\lambda \gg $ the coherence length. In the limit of $\lambda \rightarrow
\infty$, this potential reduces to the logarithmic potential studied in the previous
sections.

The spectrum for the parabolically confined system is shown in Fig~\ref{freqsvs} for
penetration lengths $\lambda=1, 10$ and 50. One can clearly see that for $\lambda=1$ no
upper bound frequency is found, while it is for $\lambda=50$. That the logarithmic
limit is not at all reached for $\lambda=1$ is also seen in Fig.~\ref{modesvs10} which
shows the eigenmode corresponding to the largest frequency for $N=10$. As shown above,
this eigenmode corresponds with the breathing mode in case of a logarithmic
interaction. This is not at all the case for $\lambda=1$ and 10 (see
Figs.\ref{modesvs10}(a) and (b)), but for $\lambda=50$ it does. Therefore, transition
to the logarithmic-type of interaction takes place between $\lambda=10$ and 50.

It is also interesting to look at the eigenmodes corresponding to the highest frequency
for $N=20$ (see Figs.~\ref{modesvs20}(a-c)). Now the modes for $\lambda=1, 10$ and 50
look all similar. However, they differ from the breathing mode: the middle and outer
ring move in the opposite direction. From the spectrum one can observe that for $N>10$
there are two highest frequencies which lie very close together. These highest but one
frequencies are therefore plot in Fig.~\ref{modesvs20}(d-f). Notice that for
$\lambda=10$ and 50 this mode indeed corresponds with the breathing mode.

\section{Conclusions}

We investigated the ground state structure and the spectral properties of a classical
2D cluster with a finite number of particles interacting through a repulsive
logarithmic potential. The logarithmic potential is the Coulomb potential in a 2D
world. It was shown that for a hard wall confinement all particles are situated on one
ring at the boundary of the potential, as a consequence of Earnshaw's theorem in 2D.
Multi-ring configurations exist for soft confinement potentials in which case only the
external confinement delivers the restoring force. The particles are uniformly
distributed for a parabolic confinement.

We also found that, independently of the number of particles in the cluster, all
eigenfrequencies lie below the frequency of a particular mode, namely the
breathing-like mode. This is again a consequence of Earnshaw's theorem because adding
extra particles does not create extra minima in the potential nor makes it more
complex. In the case of a parabolic confinement the frequency of this breathing mode is
independent from $N$, which results in an upper bound for the frequencies.

It is also shown that for vortices in a type II superconductor (for which the
penetration length $\lambda \gg$ coherence length) the transition to the
logarithmic-type of interaction takes place between $10< \lambda < 50$. It is not
always the highest mode anymore which corresponds with the breathing mode, it can also
be the highest but one mode.

\acknowledgments{This work is supported by the Flemish Science Foundation.}

\end{document}